\documentclass[aps,showpacs,floatfix,superscriptaddress,12pt]{revtex4-1}
\usepackage{times}

\usepackage{amssymb}
\usepackage[dvipdfmx]{graphicx}
\usepackage{xcolor}
\newcounter{lastnote}

\begin{document} 
% Include your paper's title here

\title{Universal trimers as building blocks of stable quantum matter}

% Place the author information here.  Please hand-code the contact
% information and notecalls; do *not* use \footnote commands.  Let the
% author contact information appear immediately below the author names
% as shown.  We would also prefer that you don't change the type-size
% settings shown here.

\author{Shimpei Endo}
\affiliation{Laboratoire Kastler-Brossel, \'{E}cole Normale Sup\'{e}rieure, 24 rue Lhomond, 75231 Paris, France}
\author{Antonio M. Garc\'{\i}a-Garc\'{\i}a}
\affiliation{University of Cambridge, Cavendish Laboratory, JJ Thomson Avenue, Cambridge, CB3 0HE, UK}
\author{Pascal Naidon}
\thanks{Author contributions: SE carried out the Born-Oppenhaimer calculation and the analysis of the experimental feasibility of the theoretical model.
AMG initiated the project and wrote a first draft of the manuscript. 
PN computed the effective trimer-trimer potential and the scattering length by the resonating group method. 
}
\affiliation{RIKEN Nishina Centre, RIKEN, Wako 351-0198, Japan}
%\\
%\normalsize{$^\ast$To whom correspondence should be addressed;}\\
%\normalsize{E-mail: pascal@riken.jp,  amg73@cam.ac.uk, shimpei.endo@lkb.ens.fr}
%}

% Include the date command, but leave its argument blank.

%\date{}

%%%%%%%%%%%%%%%%% END OF PREAMBLE %%%%%%%%%%%%%%%%


%\baselineskip24pt



\begin{abstract}
%Interactions that induce clusters are often also associated to stable quantum matter. Examples include the colour %quark interaction responsible for nucleons and nuclear matter, or electron-phonon interactions leading to %superconductivity by Cooper-pair condensation. 
We present an exploratory study that suggests that Efimov physics, a leading research theme in few-body quantum physics, can also induce stable many-body ground states whose building blocks are universal clusters. We identify a range of parameters in a mass-and-density imbalanced two-species fermionic mixture for which the ground state is a gas of Efimov-related universal trimers. An explicit calculation of the trimer-trimer interaction reveals that the trimer phase is an SU$(3)$ Fermi liquid stable against recombination losses. We propose to experimentally observe this phase in a fermionic $^6$Li-$^{53}$Cr mixture.
\end{abstract}

\maketitle 
\section{Introduction}
The Efimov effect~\cite{efimov1,efimov1973energy,ferlaino2010forty}, predicted by Vitaly Efimov more than forty years ago, states that three particles interacting through a sufficiently strong two-body interaction can occupy a tower of three-body bound states usually referred to as Efimov trimers. Intriguingly, these states can exist even if there are no two-body bound states, and feature discrete scale invariance. 
The binding mechanism is an effective three-body attraction that has no classical analogue.
% Efimov physics has been studied in several fields~\cite{halo,helium,helium1,efimovexp,efimovspec,PhysRevLett.%112.190401,PhysRevLett.113.240402,PhysRevLett.112.250404,ferlaino2010forty,PhysRevLett.%107.120401,efimovfour,magnons,PhysRevLett.111.026402}.
%Initially, applications were focused in nuclear physics~\cite{halo} and in clusters of $^4$He atoms~%\cite{helium,helium1}. 
The landmark  experimental observation of an Efimov trimer in a gas of Cesium atoms~\cite{efimovexp} provided new impetus to this problem.
% and shifted the interest to cold-atom physics, and more recently to condensed matter physics~\cite{magnons,PhysRevLett.111.026402}. 
Theoretical features such as the discrete geometric scaling of the energy levels corresponding to trimers~\cite{efimovspec,PhysRevLett.112.190401,PhysRevLett.113.240402,PhysRevLett.112.250404} and the universality of Efimov physics~\cite{ferlaino2010forty,PhysRevLett.107.120401} were observed experimentally in the vicinity of Feshbach resonances.
% Moreover, the existence of bound states of four particles, tetramers, associated with the Efimov trimers was reported in Ref.~\cite{efimovfour}.
 All these experiments were carried out in the dilute limit where many-body effects are negligible. 
 
Some effects of a many-body environment on Efimov physics have already been studied theoretically~\cite{pauli,pauli1,endo2013perfect,Nygaard2014,meera,PhysRevA.85.013609,efimanybose,nishidamulti}. Pauli blocking due to a static Fermi sea tends to suppress the Efimov trimers as density increases~\cite{pauli,pauli1,endo2013perfect,Nygaard2014}. By contrast, the presence of a single species Fermi sea enhances the stability of a single $p$-wave universal (non-Efimov) trimer composed of two species of fermions~\cite{meera}. However, the more fundamental question of whether Efimov physics can induce novel many-body phases is only starting to be addressed in the literature. The calculation of the third virial coefficient in a mass-imbalanced two-component Fermi gas~\cite{PhysRevA.85.013609} and Bose gas at unitarity~\cite{efimanybose} suggests an Efimov-driven phase transition. For a three-component Fermi gas the possible existence of a trimer phase was proposed in Ref.~\cite{nishidamulti}. However, in ultracold atom systems, the formation of Efimov trimers entails three-body recombination processes leading to a significant atom number loss~\cite{efimovexp,efimovspec,PhysRevLett.112.190401,PhysRevLett.113.240402,PhysRevLett.112.250404,ferlaino2010forty,PhysRevLett.107.120401}, and it is unclear whether it is possible to observe this phase in these systems. 

Here, we propose a system closely related to Efimov physics, where a many-body trimer phase can be stably realized. We show that in a mass-and-density imbalanced two-component Fermi gas, a strong enough interaction between the two components leads to a many-body phase composed of trimers.
% as schematically shown in Fig.~1. 
These trimers are not Efimov trimers but universal trimers, also known as the Kartavtsev-Malykh trimers~\cite{kartavtsev2007low}, which are precursors of the Efimov effect occurring at large mass ratios. We explicitly investigate the interaction between the trimers to characterize the nature of the trimer phase. We find the trimer-timer interaction is repulsive at low energy, which suggests that the trimer phase is a stable Fermi liquid.

\section{Trimers in mass and density imbalanced Fermi mixtures}

We study a homogeneous two-species Fermi mixture where the mass and number density of each component is denoted by $m_{\uparrow,\downarrow}$ and $n_{\uparrow,\downarrow}$ respectively. The interaction between different components is assumed to be attractive and near-resonant, i.e. it is characterised by a large $s$-wave scattering length $\vert a \vert \gg r_0$, where $r_0$ is the typical range of the interaction. It can thus be modeled by a zero-range contact interaction with scattering length $a$.
The inverse scattering length $1/a$ goes from $-\infty$ to $+\infty$ as the strength of the interaction increases, and is zero exactly at the resonance. 

For equal masses $m_{\uparrow}=m_{\downarrow}=m$, and equal densities $n_{\uparrow}=n_{\uparrow}=n$, the ground state evolves smoothly from a Bardeen-Cooper-Schrieffer (BCS) phase characterised by overlapping Cooper pairs in the weak-coupling limit $n^{1/3}a \to 0^-$, into a Bose-Einstein Condensate (BEC) of dimers with a binding energy $\displaystyle E_{\mathrm{B}}^{(2)} = - \hbar^2 /m a^2$ in the strong-coupling limit $n^{1/3}a \to 0^+$ ~\cite{chen2005bcs}. For $n_{\uparrow} \neq n_{\downarrow}$ the phase diagram is richer~\cite{PhysRevA.74.063628,PhysRevA.73.051603}, but the ground state is still a superposition of Cooper pairs or dimers rather than of trimers or tetramers.

By contrast, for a mass-imbalanced system $m_{\uparrow} > m_{\downarrow}$ \cite{nishida2008,efimov1973energy,PhysRevA.67.010703}, trimers and tetramers can be formed due to the Efimov attraction developing in odd angular momentum channels. For a three-body system of $\uparrow \uparrow \downarrow$ particles, Efimov trimers exist for a sufficiently large mass ratio $m_{\uparrow}/m_{\downarrow}>13.6$~\cite{efimov1973energy,PhysRevA.67.010703}. These trimers cannot be realized stably in cold-atom experiments due to their internal instability against three-body recombination~\cite{efimovexp,efimovspec,PhysRevLett.112.190401,PhysRevLett.113.240402,PhysRevLett.112.250404,ferlaino2010forty,PhysRevLett.107.120401}: at short distances of the order of $r_0$, two atoms of the trimer can recombine into a tightly bound diatomic molecule, whose binding energy is released by the third atom. On the other hand, for a mass ratio $8.2<m_{\uparrow}/m_{\downarrow}<13.6$, there exists a finite number of trimer states, the so-called universal trimers (also known as Kartavtsev-Malykh trimers \cite{kartavtsev2007low}) that carry one unit of angular momentum and have a binding energy  $E_{\mathrm{B}}^{(3)}\propto E_{\mathrm{B}}^{(2)} \propto a^{-2}$
. These states are precursors of the Efimov states occurring at larger mass ratios~\cite{kartavtsev2007low,endo2012universal}. Although they do not exhibit discrete scaling invariance, their existence originates from a continuation of the Efimov three-body attraction to mass ratios smaller than 13.6. For $8.2 <m_{\uparrow}/m_{\downarrow}<12.9$, there is only one universal trimer state, while there exist two universal trimer states for $12.9 <m_{\uparrow}/m_{\downarrow}<13.6$.  These states are triply degenerate, due to their angular momentum $\ell=1$. Unlike the Efimov trimers, the universal trimers are internally stable against decay by three-body recombination to possible deeper states \cite{kartavtsev2007low,PhysRevLett.103.153202} because the centrifugal potential between identical $\uparrow$-$\uparrow$ fermions is strong enough to prevent three particles from coming within distances of the order of $r_0$ where the three-body recombination processes can occur. 

According to Ref.~\cite{PhysRevLett.109.230404}, a universal tetramer state, composed of three heavy and one light particles, exists for $m_{\uparrow}/m_{\downarrow} \gtrsim 9.5$. Its existence is also related to the occurrence of a four-body Efimov attraction for mass ratio $m_{\uparrow}/m_{\downarrow} > 13.4$~\cite{PhysRevLett.105.223201}. The existence of larger clusters, pentamers or hexamers, is still an open question, but since the Pauli exclusion principle between the identical fermions is more significant for larger number of particles, it is reasonable to assume that they do not exist in the mass ratio range $8.2<m_{\uparrow}/m_{\downarrow}\lesssim 9.5$. Therefore, for $8.2<m_{\uparrow}/m_{\downarrow}\lesssim 9.5$, the lowest energy few-body bound state is a universal trimer. In this paper, we focus on mass ratios around this range, since our main aim is to investigate the possibility of a stable trimer phase connected to Efimov physics.

In order to have a glimpse of the region of parameters in which a gas of trimers is the ground state, let us consider the system for a general density imbalance in the limit $n a^3 \rightarrow 0^+$ limit where interactions among trimers, dimers, or atoms can be made arbitrarily weak and the ground state is simply the configuration of very weakly-interacting trimers, dimers and unpaired single particles. Let us first neglect these weak interactions, which will be studied in the following section for the trimer-trimer case, and investigate the phase diagram as a function of density imbalance. The key for this analysis is the fact that for the mass ratio $8.2<m_{\uparrow}/m_{\downarrow}\lesssim 9.5$, the binding energy of the trimers is larger than, but less than twice, that of the dimer, i.e. $\displaystyle \left|E_{\mathrm{B}}^{(2)}\right|<\left|E_{\mathrm{B}}^{(3)}\right|<2\left|E_{\mathrm{B}}^{(2)}\right|$~\cite{kartavtsev2007low,endo2011universal}. Therefore, the trimers can only appear when the number of $\uparrow$ fermions exceeds that of $\downarrow$ fermions: for $n_{\uparrow}/n_{\mathrm{tot}}<1/2$, the ground state is a Bose-Fermi mixture of dimers and $\downarrow$ fermions as shown in Fig.~2. In the region $n_{\uparrow}/n_{\mathrm{tot}} > 1/2$, the universal trimers appear, behaving as fermions since they consist of three fermions. For $\displaystyle 1/2<n_{\uparrow}/n_{\mathrm{tot}}<2/3$, where $n_{\mathrm{tot}}=n_{\uparrow}+n_{\downarrow}$, the ground state is a Bose-Fermi mixture of dimers and trimers. Since the trimer state carries one unit of angular momentum, the Fermi surface becomes triply degenerate, and can be labelled by the rotational number $m = \pm 1,0$. For a larger population imbalance $n_{\uparrow}/n_{\mathrm{tot}}>2/3$, the ground state is a mixture of heavy fermions and trimers.

At $n_{\uparrow}/n_{\mathrm{tot}} = 2/3$, there are no dimers or remaining fermions. The ground state in the dilute limit, that neglects any trimer-trimer interaction, is a three-component Fermi system of non-interacting universal trimers. It is remarkable that a three-component Fermi system emerges from the original two-species mass-imbalanced Fermi system. From now on we focus exclusively on this region $n_{\uparrow}/n_{\mathrm{tot}} \approx 2/3$.

\section{Low-energy effective Hamiltonian of the trimer phase}

Since the centrifugal repulsion between the $\uparrow$ fermions is strong enough to overcome the Efimov attraction at short distances and suppress the $3$-body and $4$-body Efimov effects in the mass ratio window we are interested in $8.2 <m_{\uparrow}/m_{\downarrow}\lesssim 9.5$~\cite{efimov1973energy,PhysRevA.67.010703,PhysRevLett.105.223201,endocastin}, we can reasonably assume that the same argument applies to $5$-body and $6$-body Efimov effects. The scattering properties between the trimers are therefore universally characterised by the $s$-wave scattering length $a$ between the $\uparrow$ and $\downarrow$ fermions. Each trimer carries one unit of angular momentum therefore three scattering channels, labelled by the total internal angular moment $F=0,1,2$ of the six-body problem, are available. Since the trimers are fermions, the relative orbital state has an even parity for $F=1$, while it has an odd parity for $F=0,2$. The $s$-wave scattering occurs only in the $F=1$ channel, while the dominant channel is the $p$-wave scattering in the $F=0,2$ channels. In the zero-temperature limit, $s$-wave scattering dominates and the effective trimer-trimer interaction is controlled by the $F=1$ channel. That implies that interactions are restricted to trimers with different projections of the angular momentum. It is important to note that the contribution from the other channels is not necessarily negligible in the range of temperatures relevant to experiments. However in line with the exploratory character of this study we focus on the zero temperature limit that is well approximated by the  following trimer-trimer interacting Hamiltonian,
\begin{eqnarray}
H_{\mathrm{int}} \approx \sum_{ij} V_{{\scriptscriptstyle F=1}}(\mbox{\boldmath $r$}_i - \mbox{\boldmath $r$}_j) P_{{\scriptscriptstyle F=1}}^{ij}, \label{effint}
\end{eqnarray}
where we are assuming that trimers are point particles at $\mbox{\boldmath $r$}_i$ and $\mbox{\boldmath $r$}_j$, $V_{{\scriptstyle F}}$ is the corresponding interaction potential, $\displaystyle P_{{\scriptscriptstyle F}}^{ij}=\sum_{M=-F}^F \left | F,M \rangle_{{\scriptscriptstyle ij}}\   {}_{{\scriptscriptstyle ij}}\langle  F,M \right|$ is the projection operator onto a two-body state of the $i$-th and $j$-th trimers in a basis given by the total internal angular momentum $F$ and its total projection $M$. We note that in realistic experimental situations the ground state may not only be composed of trimers though this is not important because it is experimentally possible to remove other particles from the system.

%In the low-energy and long-distance limit ($r \gg r_0$), the effective interaction Hamiltonian has the zero-range form:
%\begin{eqnarray}
%H^{\mathrm{(eff)}}_{\mathrm{int}} = g_{{\scriptscriptstyle F=1}} \sum_{ij} \delta^{(3)}(\mbox{\boldmath $r$}_i - %%\mbox{\boldmath $r$}_j) P_{{\scriptscriptstyle F=1}}^{ij}, \label{effintlowenergy}
%\end{eqnarray}
%where the coupling constant $\displaystyle g_{{\scriptscriptstyle F=1}} = \int d^3\mbox{\boldmath $r$} %V_{{\scriptscriptstyle F=1}}(r) \approx \frac{4 \pi \hbar^2}{2m_{\uparrow}+m_{\downarrow}} a^{t t}$ is set by the %trimer-trimer $s$-wave scattering length $a^{t t}$. 
Using the Clebsch-Gordan coefficients for $F=1$, Eq.(\ref{effint}) is rewritten in the second quantization form as 
\begin{eqnarray}
\nonumber H^{\mathrm{(eff)}}_{\mathrm{int}} &=& \int d^3\mbox{\boldmath $r$} \int d^3\mbox{\boldmath $r'$}V_{{\scriptscriptstyle F=1}}(\mbox{\boldmath $r$} - \mbox{\boldmath $r'$}) \Biggl[\psi_{1}^{\dagger} (\mbox{\boldmath $r$})\psi_{0}^{\dagger} (\mbox{\boldmath $r'$})\psi_{0} (\mbox{\boldmath $r'$})\psi_{1} (\mbox{\boldmath $r$})+ \psi_{1}^{\dagger} (\mbox{\boldmath $r$})\psi_{-1}^{\dagger} (\mbox{\boldmath $r'$})\psi_{-1} (\mbox{\boldmath $r'$})\psi_{1} (\mbox{\boldmath $r$}) \\ 
\nonumber  & & \hspace{2.5cm} + \psi_{-1}^{\dagger} (\mbox{\boldmath $r$})\psi_{0}^{\dagger} (\mbox{\boldmath $r'$})\psi_{0} (\mbox{\boldmath $r'$})\psi_{-1} (\mbox{\boldmath $r$})\Biggr]\\
\label{eq:su3_ham} & \approx & \frac{1}{2}\sum_{m_1 m_2}\int d^3\mbox{\boldmath $r$}d^3\mbox{\boldmath $r'$}V_{{\scriptscriptstyle F=1}}(\mbox{\boldmath $r$} - \mbox{\boldmath $r'$})\psi_{m_1}^{\dagger} (\mbox{\boldmath $r$})\psi_{m_2}^{\dagger} (\mbox{\boldmath $r'$})\psi_{m_2} (\mbox{\boldmath $r'$})\psi_{m_1} (\mbox{\boldmath $r$}),
\end{eqnarray}
where $m_i=\pm 1,0$ denotes the trimer quantum number related to the projection of the angular momentum. In the last line above we assume that $V_{{\scriptscriptstyle F=1}}(\mbox{\boldmath $r$} - \mbox{\boldmath $r'$})$, to be computed explicitly in the next section, is a  short-range potential.  This assumption makes the terms corresponding to $m_1 = m_2$ negligible due to Fermi statistics. 
Interestingly, the last line in Eq.~(\ref{eq:su3_ham}) is SU(3) symmetric~\cite{fn1}, even though the initial heavy-light mixture has much less symmetry. This enhanced SU(3) symmetry could be observed experimentally provided that the potential is sufficiently short-ranged and other processes that break the degeneracy of $m$-states, such as dipole-dipole interactions, are negligible. 

Finally we note that due to the short-range form of Eq.~(\ref{eq:su3_ham}), spin-flipping processes are heavily suppressed. Therefore, the initial population in each internal angular momentum state $n_{\pm 1}, n_{0}$ is an almost conserved quantity. In order for the system to evolve into an equilibrium state $n_{-1}= n_{1} = n_{0}$, it is necessary to consider spin-flipping processes occurring in the higher-partial-wave scattering channels. We shall see later that this sets a natural time scale for equilibration.

\section{Trimer-Trimer potential and the nature of the ground state}
The nature and stability of the trimer phase strongly depends on the trimer-trimer interaction potential $V_{{\scriptscriptstyle F=1}}$. If it is purely repulsive, then $a^{tt}_{{\scriptscriptstyle F=1}}>0$ and the ground state is a three-component Fermi liquid composed of trimers. If it is sufficiently attractive so that $a^{tt}_{{\scriptscriptstyle F=1}}<0$, then it leads to an SU(3) superfluid state~\cite{modawi1997some,PhysRevA.82.063615} in which the building blocks are not pairs of particles but pairs of trimers. However, if the attraction is too strong, this trimer phase can be unstable against collisional losses in ultracold atom experiments, which occur when the atoms of colliding trimers come within distances of the order of $r_0$ and recombine into more deeply bound molecules. 

To understand better the trimer phase, we need to carry out an explicit calculation of the trimer-trimer potential and the associated $s$-wave scattering length in the $F=1$ channel. This is reminiscent of the problem of the baryon-baryon potential~\cite{oka,detar,detar1} in nuclear physics within the quark model. In both problems, basic constituents are fermions, so the Pauli principle is expected to play an important role especially for short distances. However, there are important differences. For instance, in the universal trimer case, a long-range attraction between the heavy $\uparrow$ fermions can be induced by exchanging the light $\downarrow$ fermions between the trimers. This is in marked contrast to the short-range character of typical nuclear forces. The determination of the trimer-trimer interaction is a rather non-trivial problem, requiring a six-body calculation. Solving exactly the six-body problem, however, is a very demanding numerical calculation, so we opt here for approximate methods. 

We first compute the trimer-trimer potential by the Born-Oppenheimer (BO) approximation. The three-body physics, including the appearance of the universal trimers and the Efimov trimers, can be adequately reproduced by the BO approximation thanks to the large mass imbalance. It is therefore natural to extend the BO approximation to the six-body problem as a simple but heuristic approach to understand the trimer-trimer scattering process. In the BO approximation, the light $\downarrow$ fermions' Schr\"odinger equation is solved by fixing the heavy $\uparrow$ particles' positions. The eigenenergy of the light fermions acts as an adiabatic potential between the $\uparrow$ particles. Subtracting the internal energy of each trimer, assuming the two pairs of heavy particles are well separated, we obtain the trimer-trimer potential (see Supplementary information for more details \cite{supplemental}). In Figs.~3 (a) and (b), we show the trimer-trimer potential calculated with the BO method for various heavy particles' configurations sketched in Figs.~3 (c) to (h). The trimer-trimer potential is repulsive at large distance for all configurations, decaying exponentially as the dimer wave function of the light fermion around the heavy particle. The repulsion is due in part to the fermionic nature of the light particles: although the first light fermion occupies a bonding orbital, the second light fermion has to occupy an antibonding orbital which overcomes the attraction of the bonding orbital. We note that the centrifugal repulsion between the heavy fermions are not included in the BO potential shown in Fig.~3. Therefore, the actual trimer-trimer interaction would be more repulsive than that shown in Fig.~3. 

The BO approach has a several shortcomings: it is not realistic to compute the trimer-trimer scattering length as it requires solving a $4$-body Schr\"odinger equation with the long-range BO potential, which would be hardly easier than solving the original six-body problem with a contact interaction. Moreover, it cannot correctly capture the trimer-trimer potential at short distance where, see Fig.~3, the potential is attractive for most of the configurations. We note that the attraction is likely an artefact of the way in which we subtract the trimer internal energy, which is strictly valid only when the two trimers are well separated (see Supplementary information \cite{supplemental}), and the fact that the centrifugal repulsion among heavy fermions is not included.  In summary, the BO results depicted in Fig.~3 could be interpreted as the trimer-trimer potential only when the trimer-timer separation is much larger than the trimer size $R\gtrsim a$. 

Faced with these limitations, we use a second approximate method to evaluate the trimer-trimer potential and the scattering length that addresses satisfactorily some of these problems: the resonating group method (RGM)~\cite{rgm,rgm1,rgm2,rgm3} commonly employed in nuclear physics calculations. Unlike the BO method, the RGM takes into account exactly the antisymmetrization of the wavefunctions due to the Pauli principle which typically plays an important role for short distances. The RGM approximates the six-body wave function describing the trimer-trimer scattering by the antisymmetrized product of the known trimer wave functions and an unknown wave function describing the relative motion between the two trimers. Therefore, it is a good approximation provided that, except for the exchange of identical particles, the wave functions for the three-body subsystems are not substantially altered during the scattering process. More specifically, the RGM approach is qualitatively correct unless the trimer-trimer interaction is strongly attractive so that there is a six-body bound state. Although we cannot rule out that this is the case, the fact that a tetramer state only appears at a larger mass ratio \cite{PhysRevLett.109.230404} than the ground state trimer suggests that, a six-body bound state, if exists, it must occur at an even larger mass ratio.

From the RGM ansatz, one can derive an effective interaction potential between the two trimers~\cite{Naidon1}. As is shown in Fig.~4 (a), we have found that this potential is repulsive for all distances. For large separations, the decay of the potential is exponential, in agreement with the BO results. In Fig.~4 (b), we present the trimer-trimer scattering length for different mass imbalances obtained from the RGM trimer-trimer potentials. For all mass ratios between 8.2
and 12.9, for which only one trimer state exists, the scattering length is found to be positive.

Although not exact, the RGM method indicates that, due to Fermi statistics, trimers experience a repulsion for separations of the order of the scattering length and larger. The RGM prediction that the interaction is still repulsive at smaller separations is valid provided that the trimer-trimer scattering does not result in a six-body bound state in the range of mass ratios of interest.  This strongly suggests the possibility that colliding trimers, like dimers~\cite{PhysRevLett.93.090404,PhysRevLett.93.050401}, are protected from collisional losses occurring at shorter distances, raising the prospect of the experimental realization of the trimer phase in ultracold-atom experiments. As to the nature of the trimer phase, the positive trimer-trimer scattering length obtained in our RGM calculation indicates that it should be an SU$(3)$ Fermi liquid, rather than an SU$(3)$ superfluid. The RGM calculation being approximate, we cannot completely rule out the possibility of the latter. Indeed, we expect that an important correction, that increases with the mass ratio, comes from the virtual excitations of the trimers into the dimer-particle continuum. Perturbation theory to second order indicates that this correction is attractive. An explicit calculation would be required to clarify whether, for certain mass ratios, the strength of this correction is strong enough to destabilize the Fermi liquid ground state. The most interesting scenario would correspond to a small attraction at large distance inducing superfluidity, while preserving the stabilizing Fermi repulsion at short distance.
\section{Experimental realization}
Prime candidates to observe experimentally this novel trimer phase are fermion mixtures of ultracold atoms. A mixture of $^6$Li and $^{53}$Cr atoms~\cite{PhysRevA.91.011603} ($m_{\uparrow}/m_{\downarrow}=8.80$) falls within the mass-ratio window $8.2<m_{\uparrow}/m_{\downarrow}\lesssim 9.5$~\cite{fn2} in which a trimer Fermi liquid phase is possible. If the $s$-wave scattering length is tuned to satisfy the condition $\lambda,  n_{\mathrm{tot}}^{-1/3} \gg a \gg r_{\mathrm{vdw}}$ via a Feshbach resonance, where $r_{\mathrm{vdw}}$ is the van der Waals length of the atoms, the trimer Fermi liquid phase can be realized at the density  imbalance $n_{\uparrow}/n_{\mathrm{tot}} \approx 2/3$. Another experimental setting of interest is a Li-K mixture where radio-frequency techniques have shown\cite{jag2014} a strong atom-dimer p-wave attraction at a lower mass ratio of $6.63$.

In any case a necessary condition for the observation of the trimer phase is that the typical time scale for its formation and equilibration is within the reach for current cold-atom experiments. Without any external influence, the trimer phase forms via a four-body recombination of the fermionic mixture, which is too slow to be observed in cold-atom experiments. However, trimers can be formed within a much shorter time compatible with experimental observation by using an adiabatic change of the $s$-wave scattering length~\cite{RevModPhys.78.1311}, or performing the radio-frequency association of trimers in an atom-dimer mixture~\cite{PhysRevLett.106.143201,lompe2010radio}. Once the trimers are formed by these methods, one must wait for the trimer gas to equilibrate. There are two important time scales for the  equilibration: thermalization time $\tau_{\mathrm{th}}$ and relaxation time of the trimer's internal states $\tau_{\mathrm{int}}$. Since the trimers can change their internal states only via the $p$-wave collision, the relaxation time $\tau_{\mathrm{rel}}\sim \hbar(k_F^{(\mathrm{tot})} a)^{-6}/E_F^{(\mathrm{tot})} $, where $k_F^{(\mathrm{tot})}$ and $E_F^{(\mathrm{tot})}$ are the total Fermi wavevector and Fermi energy of the mixture respectively, is much larger than the thermalisation time $\tau_{\mathrm{th}}\sim \hbar(k_F^{(\mathrm{tot})} a)^{-2}/E_F^{(\mathrm{tot})} $, which occurs via $s$-wave scattering. Therefore, the relaxation time gives the minimum required time in experiments to realize an equilibrium trimer Fermi liquid phase. For $a \sim 100$nm ,  $k_F^{(\mathrm{tot})} \sim 0.003$nm$^{-1}$,  corresponding to a density $n \sim 10^{18}$ atom$/m^3$, $E_F^{(\mathrm{tot})} = \hbar^2 {k_F^{(\mathrm{tot})}}^2/2m$, with $m$ here the mass of the atom, we obtain $k_F^{(\mathrm{tot})} a \sim 0.3$, $
\tau_{\mathrm{th}}\sim 0.1$ ms and $\tau_{\mathrm{rel}} \sim 10$ ms, which is an accessible time scale in cold-atom experiments.

Another potential source of deviations between theory and experiments is the presence of magnetic dipole interactions in the mixture that break the $SU(3)$ symmetry of the model. This effect will be noticeable only if the energy related to the magnetic dipole interaction $\mathrm{E_D} \sim \mu^2\mu_0/4\pi r^3$ is comparable with the trimer bound energy $\mathrm{E_B} \sim \hbar^2/2m^2a^2$. Assuming $r \sim a$, as the typical size of the trimer, $\mu_0$ the vacuum permeability and $\mu \approx 6\mu_B$ with $\mu_B$ the Bohr's magneton \cite{lahaye2009} it turns out that, for any $a\gg 1$nm, a condition easily met experimentally,  the splitting energy $\mathrm{E_B}$ is always much smaller than $\mathrm{E_D}$. This is a strong indication that the $SU(3)$ symmetry is to be approximately  conserved in realistic experimental conditions.

\section{Conclusion}
 We have identified a region of parameters in a mass-and-density-imbalanced two-species fermionic mixture for which the system undergoes a transition to a weakly-repulsive $SU(3)$ Fermi liquid composed of stable universal trimers. 
 %connected to the Efimov effect. 
 %The trimer-trimer interactions calculated by the Born-Oppenheimer method and the resonating group method are found to be %repulsive at low energy. This suggests that the trimer phase is a stable SU$(3)$ Fermi liquid. 
 The parameters required to realize this many-body phase are within reach with cold-atom experiments on mass-imbalanced fermionic mixtures, such as $^6$Li and $^{53}$Cr atoms. 

The present work also raises many interesting questions about the role of Efimov physics in many-body systems. What happens in the resonantly-interacting regime $n_{\mathrm{tot}} a^3 \rightarrow \infty$ ? In which circumstances can Efimov physics induce superfluidity? Is the ground state for larger mass ratio $m_{\uparrow}/m_{\downarrow}\gtrsim 9.5$ a Bose gas of universal tetramers~\cite{PhysRevLett.109.230404}? How do the trimer and tetramer phases evolve with mass ratio and population imbalance? 
The answers to these questions would be an important advance in the understanding of the conditions for Efimov-related clusters to constitute the building blocks of novel phases of quantum matter.
%\vspace{0.5cm}
%~\cite{acknow}

\vspace{0.3cm}

\acknowledgments
We are grateful to T. Hatsuda, and M. Ueda for helpful discussions. A.M.G. was supported by EPSRC, grant No.
  EP/I004637/1, FCT, grant PTDC/FIS/111348/2009 and a Marie Curie International
  Reintegration Grant PIRG07-GA-2010-268172. S.E.
  acknowledges support from JSPS. P. N. acknowledges support from RIKEN through
  the Incentive Research Project funding. 
%\bibliographystyle{Science}
%\bibliography{ref}

\begin{thebibliography}{100}
\bibitem{fn1}
The Hamiltonian formally has a global U($3$)= SU$(3)\times $U$(1) \times
  \mathbb{Z}_2$ symmetry, but here we are not interested in the extra U$(1)$
  global symmetry, related to the identity transformation, nor in the
  $\mathbb{Z}_2$ parity symmetry, so in the following we will simply refer to
  the symmetry as SU($3$).
\bibitem{fn2}
$^{53}$Cr has a large magnetic dipole moment~\cite{PhysRevA.91.011603}. The
  dipole interaction between $^{53}$Cr atoms can split the three-fold
  degeneracy of the universal trimer; $m=0$ state has lower energy than $m=\pm
  1$, breaking the SU$(3)$ symmetry. This effect is negligible and the
  Hamiltonian Eq.~(\ref{eq:su3_ham}) is valid only when the dipole interaction
  energy scale is much smaller than the other: $\displaystyle \lambda \gg
  n^{-1/3} \gg a \gg \ell_d$, where $\ell_d$ is the dipole length.
\end{thebibliography}

\clearpage

%\begin{figure}[h]
%\includegraphics[width=0.4\columnwidth,trim={0.0cm 0cm 2.0cm 0cm},clip,angle=0]{TrimerFigures01d.eps}
%\includegraphics[width=0.55\columnwidth,trim={1.0cm 1cm -1.3cm 1cm},clip,angle=0]{TrimerFigures02d.eps}
%\caption{Schematic illustration of the ground state of a mixture of light and heavy fermions, for a mass ratio in the %range $8.2 <m_{\uparrow}/m_{\downarrow}\lesssim 9.5$ and density ratio $n_{\uparrow}/n_{\mathrm{tot}} %%\sim 2/3$. a) Weak-coupling regime $n_{\mathrm{tot}} a^3 \rightarrow 0^{-}$, where light and heavy fermions are %nearly free. b) Strong-coupling regime $n_{\mathrm{tot}} a^3 \rightarrow 0^{+}$, where the light and heavy %fermions group into universal trimers. This trimer phase has an SU($3$) symmetry, originating from the triple %degeneracy of the trimer state, which has an $\ell = 1$ internal angular momentum and odd parity. The trimer-%trimer interaction, computed by the Born-Oppenheimer method (BO) and resonating group method (RGM), is %found to be repulsive. This suggests that the trimer phase is a stable SU(3) Fermi liquid.}
%\end{figure}

\begin{figure}[t]
 \includegraphics[width=0.6\textwidth,clip,angle=0]{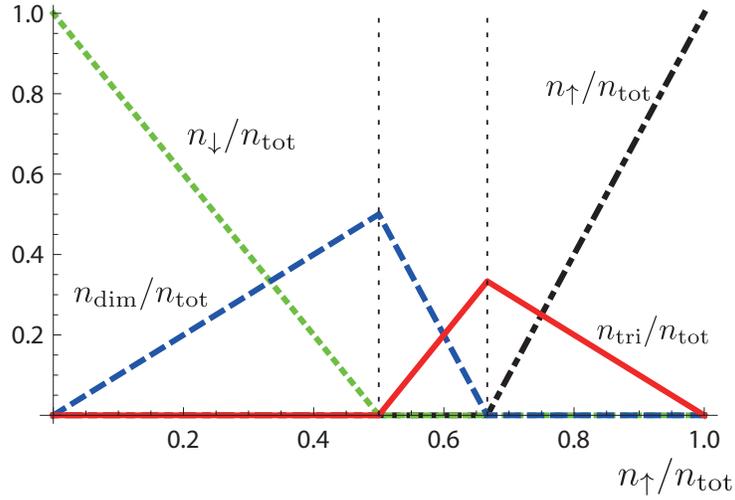}
 \caption{The number of $\uparrow$ ($n_{\uparrow})$ and $\downarrow$ ($n_{\downarrow}$) fermions, dimers $n_{\mathrm{dim}}$ and trimers $n_{\mathrm{tri}}$ as a function of the population imbalance $n_{\uparrow}/n_{\mathrm{tot}}$, where $n_{\mathrm{tot}}=n_{\uparrow}+n_{\downarrow}$ is the total fermion density. These results are strictly valid in the dilute limit that neglects completely interactions among fermions, dimers and trimers. Moreover for fermionic ground states we also neglect the effect of a finite Fermi pressure.
In this paper, we focus in the region around $n_{\uparrow}/n_{\mathrm{tot}} = 2/3$ where the ground state is a superposition of universal trimers.}  
 \end{figure}

\begin{figure*}
	\includegraphics[width=17cm,clip,angle=0]{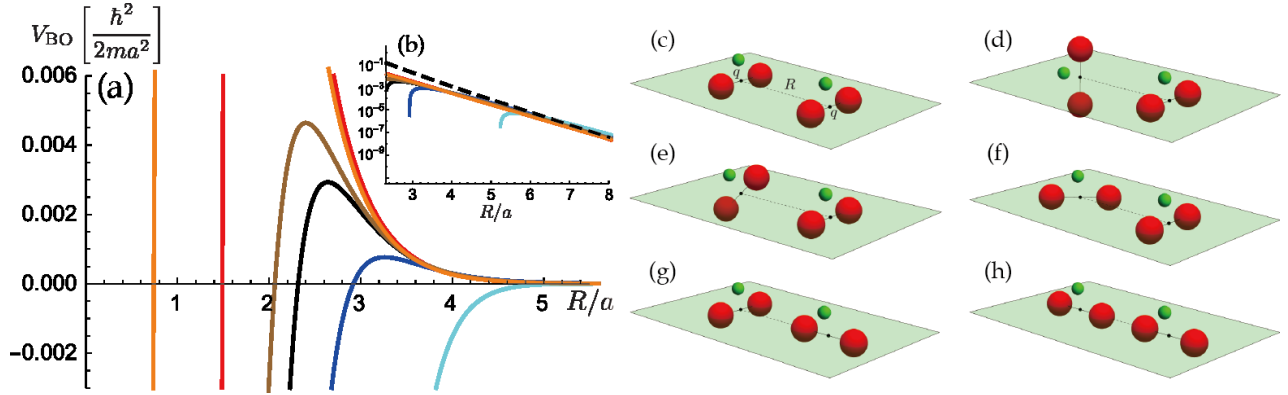}  
	\caption{(a) (b) Trimer-trimer potential, in units of $\hbar / 2ma^2$ with $m$ the mass of the light particle, calculated with the Born-Oppenheimer approximation for various configurations illustrated in (c)-(h). From (c) top (in the large $R/a$ limit) red curve to (h) bottom cyan (light grey in a greyscale) curve. The distance between the heavy particles is taken to be $q=1.6 a$. The trimer-trimer interaction is repulsive at large distance for all configurations. It decays exponentially with a decay constant very similar to that of the square of the dimer wave function (black dashed line in Fig. (b)) corresponding to a light particle around a heavy particle. }
\end{figure*}

 \begin{figure*}[t]
 	\includegraphics[width=\textwidth,clip,angle=0]{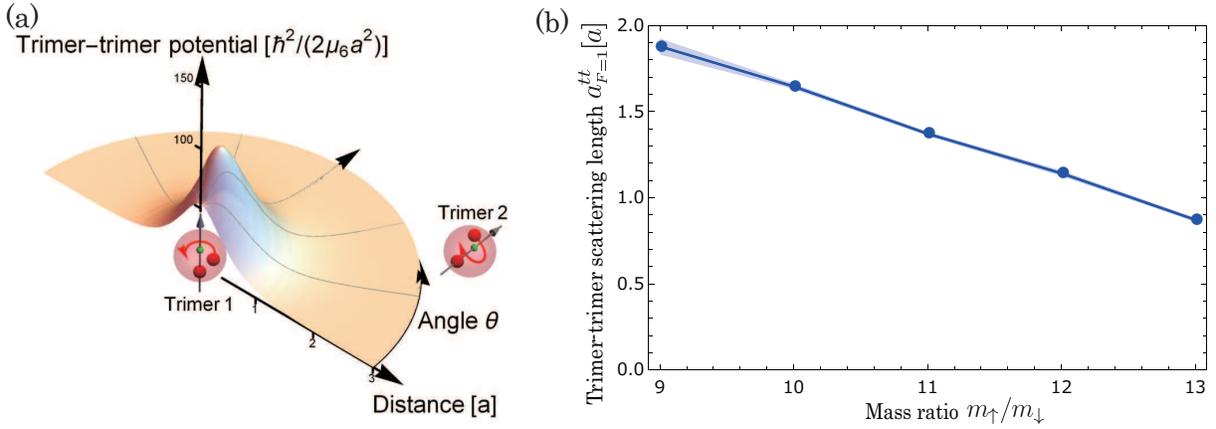}
 	\caption{(a) Trimer-trimer potential for $m_{\uparrow}/m_{\downarrow}=9$ calculated with the resonating group method, where $\mu_6=m_{\downarrow}/2+m_{\uparrow}$ is the reduced mass between the trimers. It is repulsive for any trimer-trimer separation $R$ and azimuthal angle $\theta$ between the trimers measured from the quantization axis of the trimer internal states. (b) Trimer-trimer scattering length $a^{t t}_{\scriptscriptstyle F=1}$ as a function of mass ratio calculated with the RGM method. It is always positive and decreases as the mass ratio increases. This is expected as the trimers are more tightly bound for larger mass ratio and therefore their cross section is smaller. We note that, strictly speaking, our results are only applicable for mass ratios $\leq 12.9$. For larger mass ratios an additional universal trimer state becomes available and the calculation of the scattering length becomes much more involved.}
 \end{figure*}

\end{document}